\documentclass[runningheads]{llncs}
\usepackage{graphicx}
\usepackage{multirow}
\usepackage{hyperref}
\usepackage{diagbox}
\usepackage{colortbl}
\usepackage{hhline}
\begin{document}
\title{Automatic Myocardial Disease Prediction From Delayed-Enhancement Cardiac MRI and Clinical Information} 
\titlerunning{Automatic myocardial disease prediction}

\author{Ana Lourenço\inst{1,2}
, Eric Kerfoot\inst{1}
, Irina Grigorescu\inst{1}
, Cian M Scannell\inst{1}
, Marta Varela*\inst{1,3}
, Teresa M Correia\thanks{Contributed equally.}\inst{1}
}
%
\authorrunning{A. Lourenço et al.} 
%

%
\institute{School of Biomedical Engineering and Imaging Sciences, King's College London, London, UK \and
Faculty of Sciences, University of Lisbon, Lisbon, Portugal \and  National Heart and Lung Institute, Imperial College London, London, UK \email{teresa.correia@kcl.ac.uk; marta.varela@imperial.ac.uk}}


\maketitle              
\begin{abstract}
Delayed-enhancement cardiac magnetic resonance (DE-CMR) provides important diagnostic and prognostic information on myocardial viability. The presence and extent of late gadolinium enhancement (LGE) in DE-CMR 
is negatively associated with the probability of improvement in left ventricular function after revascularization. Moreover, LGE findings can support the diagnosis of several other cardiomyopathies, but its absence does not rule them out, making disease classification by visual assessment difficult. In this work, we propose deep learning neural networks that can automatically predict myocardial disease from patient clinical information and DE-CMR. All the proposed networks achieve very good classification accuracy ($>$85\%). Including information from DE-CMR (directly as images or as metadata following DE-CMR segmentation) is valuable in this classification task, improving the accuracy to 95-100\%.

\keywords{Cardiac MRI \and Late Gadolinium Enhancement \and 
Myocardial Infarction \and Classification}
\end{abstract}

\section{Introduction}

Delayed-enhancement cardiac magnetic resonance (DE-CMR) is considered the non-invasive gold standard for assessing myocardial infarction and viability in coronary artery disease \cite{Arai_2011,Bettencourt_2009,Weinsaft_2007} and can help differentiate ischemic from non-ischemic myocardial diseases \cite{Kramer_2015}. DE-CMR images are typically acquired 10-15 minutes after an intravenous injection of a gadolinium-based contrast agent. The contrast agent is washed-out by normal tissue, but in regions with scar or fibrotic tissue, contrast washout is delayed, making nonviable regions appear bright in $T_1$-weighted images. 
The presence and extent of late gadolinium enhancement (LGE) within the left ventricular (LV) myocardium provides important diagnostic and prognostic information, including the risk of an adverse cardiac event and response to therapeutic strategies such as revascularization \cite{Kim_2000,Gerber_2012,Allman_2002,Bonow_2011}. Moreover, dark no-reflow regions (or microvascular obstruction) have been associated with worse clinical outcomes. 

Absence of LGE does not, however, rule out the presence of myocardial disease, since patients with, for example, extensive hibernating myocardium, hypertrophic cardiomyopathy, sarcoidosis or myocarditis may not show contrast uptake \cite{Soriano_2005,Kramer_2015,Lee_2020}. This makes disease classification from DE-CMR a complex task. Therefore, DE-CMR is often combined with other CMR sequences, such as $T_1$ and $T_2$ maps, to better characterize myocardial tissue alterations in various cardiomyopathies \cite{Kramer_2015}. 

Machine learning classification algorithms, such as support vector machines \cite{Narula_2016}, random forests \cite{Baebler_2015} and K-nearest neighbour \cite{Mantilla_2015}, have been used to predict the presence/absence of cardiovascular disease. These techniques require, however, complex feature extraction procedures and domain expertise to create good inputs for the classifier. On the other hand, deep learning architectures have the ability to learn features directly from the data and hence reduce the need for domain expertise and dedicated feature extraction \cite{Leiner_2019}.

In this work, we propose fully automatic neural networks (NNs) that perform binary classification for predicting normal vs pathological cases considering: 1) patient clinical information only (Clinic-NET), 2) clinical information and DE-CMR images (DOC-NET). We additionally considered whether including text-based information from independent DE-CMR segmentations could aid the classification task (Clinic-NET+ and DOC-NET+).

\section{Methods}
\subsubsection*{Clinical Images and Metadata}

The networks were trained and tested on the EMIDEC STACOM 2020 challenge dataset \cite{Lalande2020}, comprising 100 cases: 33 cases with normal CMR and 67 pathological cases. For each case, Phase Sensitive Inversion Recovery (PSIR) DE-CMR images (consisting of 5-10 short-axis (SA) slices covering the left ventricle) and 12 clinical discrete or continuous variables were provided. Clinical information included sex, age, tobacco (Y/N/Former smoker), overweight (BMI $>$ 25), arterial hypertension (Y/N), diabetes (Y/N), familial history of coronary artery disease (Y/N), ECG (ST+ (STEMI) or not), troponin (value), Killip max (1 - 4), ejection fraction of the left ventricle from echography (value), and NTproBNP (value). More details can be found in \cite{emidec,Lalande2020}. 


\subsubsection*{Image Preprocessing}
DE-CMR images had variable dimensions and were zero padded along the z direction, when necessary, to obtain 10 slices. To remove anatomical structures unrelated to the left ventricle, they were further cropped in plane to a matrix of 128 x 128, whose centre was the centroid of the the LV blood pool segmentation label. 

In the absence of ground truth segmentation labels, we propose a NN-based method to automatically perform image cropping and segmentation, as detailed below.

\subsubsection{DE-CMR segmentation}
A two-step approach based on NNs is proposed to automatically segment DE-CMR images into 3 classes (LV, healthy myocardium and LGE uptake area) and extract their volumes as additional inputs to Clinic-NET+ and DOC-NET+. These NNs are based on the 2D U-Net architecture ~\cite{Kerfoot2018,loureno2020left} and were trained separately on the EMIDEC dataset. The first NN was trained with the Dice loss function to identify the LV centre by segmenting the LV blood pool region and calculating the LV centroid coordinates. Then, the cropped images were sent to a second NN, which was trained with the generalized Dice loss function \cite{sudre2017} for LV, normal myocardium and scar segmentation. 

\subsubsection{Data augmentation}
A number of randomly-chosen data augmentation functions was applied to each DE-CMR volume. These replicate some of the expected variation in real images without modifying the cardiac features of interest and include: 1) image rotations and flips; 2) additional cropping of the images; 3) additive stochastic noise; 4) k-space corruption; 5) smooth non-rigid deformations using free-form deformations and 6) intensity and contrast scaling.

\subsubsection*{Neural Networks for myocardial disease prediction}

Two classification methods are proposed: 1) Clinic-NET, classification based on clinical information only and 2) DOC-NET, classification based on DE-CMR and Other Clinical information. As explained below, both of these NNs are further compared to other two networks that use information from previously segmented DE-CMR as further metadata inputs: Clinic-NET+ and DOC-NET+. 

The classification networks were trained using a cross-entropy loss function and the Adam optimizer with a learning rate of $0.00001$. We randomly divided the provided 100 cases into 3 datasets for: training (70 cases), validation (10 cases) and test (20 cases). Training was performed for 170 iterations. These hyperparameters were chosen after careful empirical tests. 

To assess the quality of each network, we calculated the classification's accuracy, specificity and sensitivity on the 20-case test dataset.

\subsubsection{Clinic-NET}
Clinic-NET takes the 12 provided metadata variables as inputs to a classification NN with 3 fully connected (fc) layers, which sequentially encode information in fc1 = 20, fc2 = 30, fc3 = 10 and 2 units, as shown in Figure \ref{fig:network}b. Parametric Rectified Linear Units (PReLu) are applied to the outputs of the first three layers. 

\subsubsection{DOC-NET}
\begin{figure}[t]
  \includegraphics[width=\linewidth]{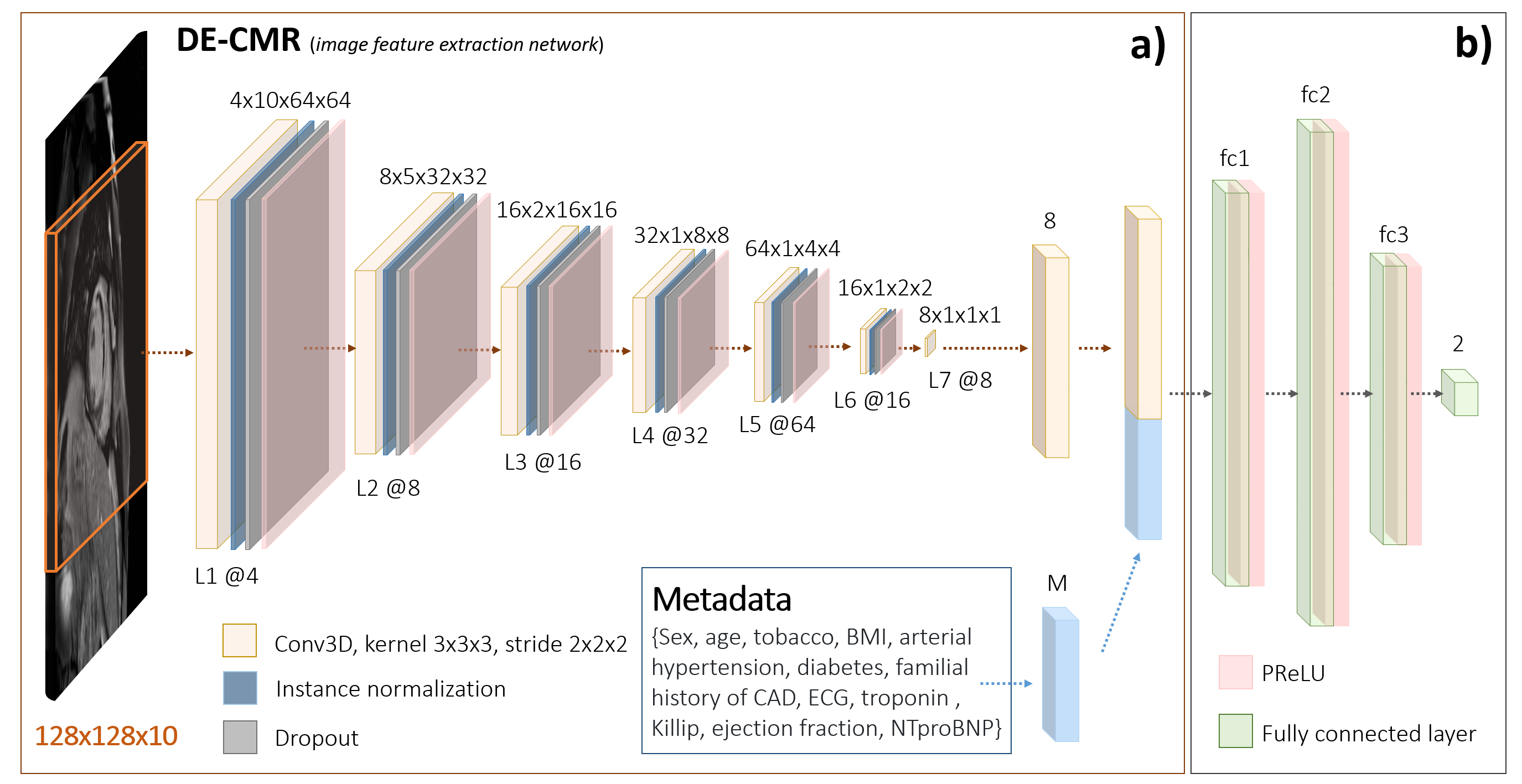}
  \caption{\textbf{DOC-NET classification network}: a) image feature vectors, obtained from the last convolutional layer of an image feature extraction network, are concatenated with the metadata vector (M) and b) sent through four fully connected (fc) layers. 
  }
  \label{fig:network}
\end{figure}

DOC-NET combines features extracted from DE-CMR images and features calculated from metadata to perform the final classification. The image feature extraction network consisted of seven layers: 1) 3D convolutions with 3x3x3 kernels, a stride of 2 and a variable number of channels (4, 8, 16, 32, 64, 16, 8); 2) instance layer-normalization; 3) dropout (20$\%$ probability of being dropped out) and 4) PReLU activation (see Fig \ref{fig:network}a). The image feature vector was then flattened into an 8-element array and concatenated with the 12-variable metadata. This combined vector was then the input to a fully connected NN similar to Clinic-Net (see Fig. \ref{fig:network}b). The sizes of the 3 fully connected layers in DOC-NET were rescaled to match the new input size, such that fc1 = 33, fc2 = 50, fc3 = 16. 

\subsubsection{Clinic-NET+ and DOC-NET+}
To further explore the classification task, we created additional metadata variables with the volumes of each of the segmentation labels of the DE-CMR. These variables were concatenated with the existing metadata and used as enhanced metadata inputs to the previously described networks to create Clinic-NET+ and DOC-NET+. For this, we used the volumes of the labels provided by the ground truth segmentations: 1) LV blood pool, 2) healthy myocardium, 3) LGE uptake area and 4) no-reflow area.

We performed additional experiments to gauge whether Clinic-Net+ and DOC-NET+ could still be deployed in more general circumstances in which expert manual segmentations are not available. In these experiments, we used the segmentation networks detailed above to automatically segment DE-CMR into labels 1-3 and used the volume of each of these categories as enhanced metadata for the classification NNs.

\begin{table}[t]
\centering
  \includegraphics[scale=0.4]{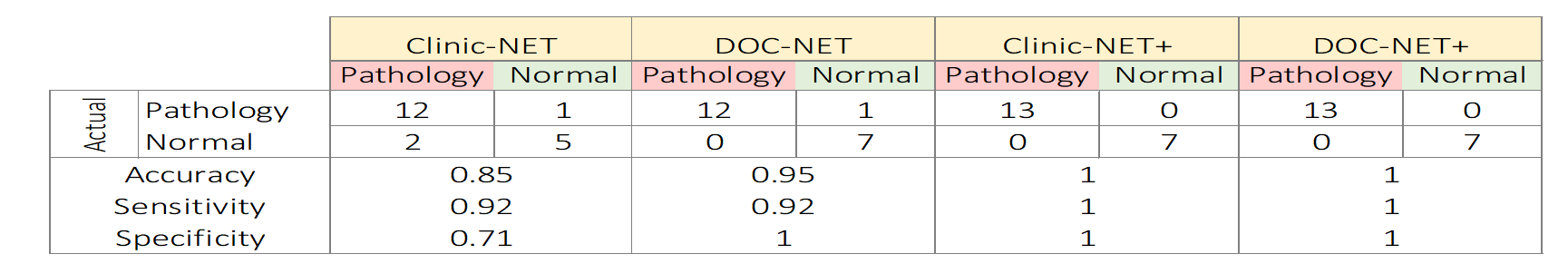}
  \caption{\textbf{Confusion matrix} (actual vs predicted counts of pathological and normal cases), \textbf{accuracy, sensitivity and specificity} obtained with the different classification networks. The additional metadata used in Clinic-NET+ and DOC-NET+ was extracted from the ground truth segmentations.}
  \label{fig:accuracy}
\end{table}

\section{Results \& Discussion}
The best overall performance was jointly achieved by Clinic-NET+ and DOC-NET+, both with an accuracy of 100\%, followed by DOC-NET (accuracy: 95\%) and Clinic-NET (accuracy: 85\%) - see Table \ref{fig:accuracy}. Our results suggest that the clinical metadata information already includes very valuable information that can be leveraged by our proposed network, Clinic-NET, to classify subjects with an accuracy of 85\%. The accuracy is greatly increased, however, when information from DE-CMR is also provided to the network. 

DOC-NET+ and Clinic-NET+ both rely on information from existing high-quality segmentations of DE-CMR performed manually by an expert or automatically by a suitable segmentation approach, such as the one proposed here (or from the EMIDEC DE-CMR segmentation challenge). We found that information about the size of potential infarct areas (and also LV dimensions) is most useful for the classification task. The excellent performance of Clinic-NET+ and DOC-Net+ is likely due to the very high predictive value of the LGE zone segmentation label, which was not present in any normal cases. 
We also investigated how the performance of Clinic-NET+ and DOC-NET+ was affected by using volumetric information from our proposed automatic segmentation method, which did not segment the no-reflow area (label 4). The classification performance was not affected, maintaining a 100\% accuracy, as can be inferred from comparing the last two columns of Table \ref{fig:accuracy} with Table \ref{fig:accuracyseg}.

\begin{figure}[t]
  \includegraphics[width=\linewidth]{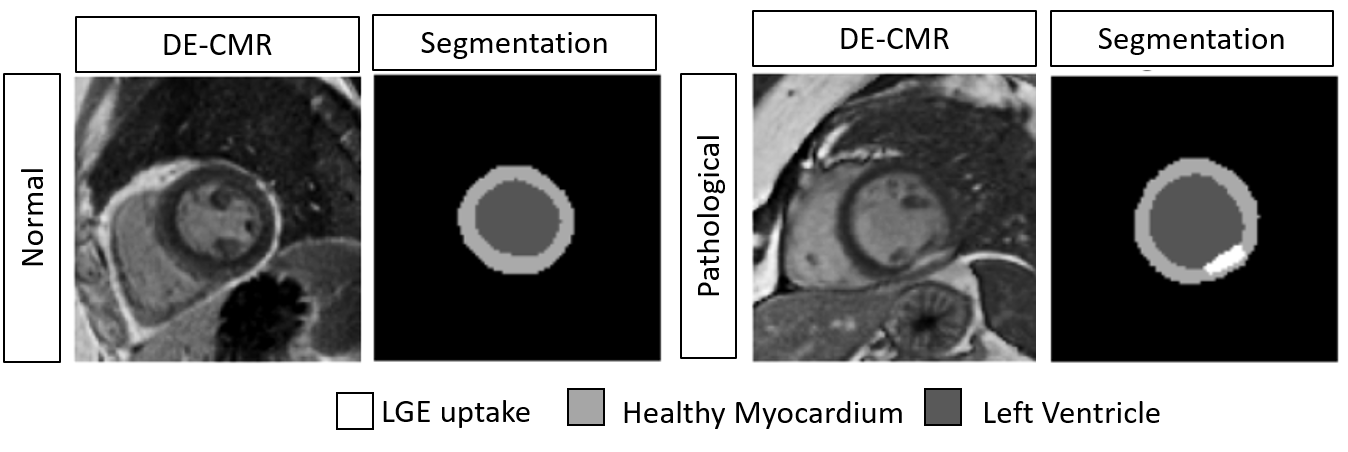}
  \caption{\textbf{DE-CMR images and segmentations} of the left ventricle, normal myocardium and region of LGE uptake (if present) obtained with the proposed automatic segmentation method for two slices from two representative subjects. 
  }
  \label{fig:seg}
\end{figure}

The proposed DE-CMR segmentation method can be particularly useful when ground truth segmentations are not available, allowing to automatically crop the region of interest and determine LV, healthy myocardium and LGE enhancement volumes (Fig. \ref{fig:seg}). However, currently, the proposed method does not segment the no-reflow area. For this particular classification task, the absence of this information did not affect the accuracy of the results. However, including information about the presence and/or volume of the no-reflow area in classification NNs may be particularly useful when predicting clinical outcomes in patients with known or potential heart problems.


The excellent results obtained when incorporating information from DE-CMR segmentations suggest that the performance of the image feature extractor included in DOC-NET/DOC-NET+ may be further enhanced when its weights are initialised with those from a well-trained segmentation network.


\begin{table}[h]
\centering
  \includegraphics[scale=0.5]{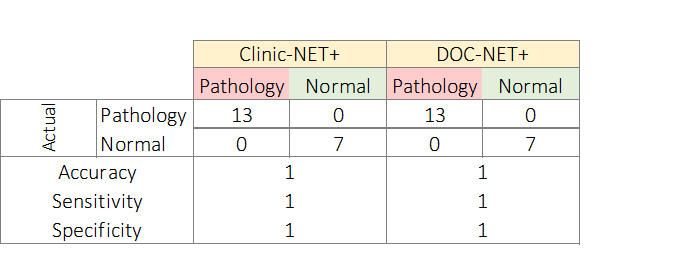}
  \caption{\textbf{Confusion matrix} (actual vs predicted counts of pathological and normal cases), \textbf{accuracy, sensitivity and specificity} obtained with the Clinic-NET+ and DOC-NET+ classification networks. The additional metadata was extracted from the automatic segmentation of DE-CMR images.}
    \vspace{-0.8cm}
  \label{fig:accuracyseg}
\end{table}

Our results were calculated in a very small dataset (20 test cases) and will be validated in a larger number of cases in the future, including in patients with undetected myocardial infarction on DE-MRI.

\section{Conclusions}
For the EMIDEC classification challenge, we propose Clinic-NET, a 4-layer fully-connected NN which uses 12 clinical variables as an input and shows a very good classification performance. An improved performance is obtained with DOC-NET, which additionally includes DE-CMR images as inputs, which are processed using 3D convolutional layers. Further performance improvements can be obtained when providing DE-CMR information distilled as the volume of segmentation labels, either from expert manual segmentations or from a proposed segmentation NN.

\section*{Acknowledgments}

This work was supported by the Wellcome/EPSRC Centre for Medical Engineering [WT 203148/Z/16/Z] and the British Heart Foundation Centre of Research Excellence at Imperial College London [RE/18/4/34215].

\bibliography{emidec.bib}

\begin{thebibliography}{10}
\providecommand{\url}[1]{\texttt{#1}}
\providecommand{\urlprefix}{URL }
\providecommand{\doi}[1]{https://doi.org/#1}

\bibitem{emidec}
{EMIDEC} classification challenge: http:/emidec.com/classification-contest
  (2020) (2020)

\bibitem{Allman_2002}
Allman, K., et~al: Myocardial viability testing and impact of revascularization
  on prognosis in patients with coronary artery disease and left ventricular
  dysfunction: a meta-analysis. J Am Coll Cardiol  \textbf{39}(7),  1151--8
  (2002)

\bibitem{Arai_2011}
Arai, A.: The cardiac magnetic resonance ({CMR}) approach to assessing
  myocardial viability. J Nucl Cardiol  \textbf{18}(6),  1095--1102 (2011)

\bibitem{Baebler_2015}
Bae{\ss}ler, B., et~al: Mapping tissue inhomogeneity in acute myocarditis: a
  novel analytical approach to quantitative myocardial edema imaging by
  {T2}-mapping. J Cardiovasc Magn Reson  \textbf{17}(1), ~115 (2015)

\bibitem{Bettencourt_2009}
Bettencourt, N., Chiribiri, A., Schuster, A., Nagel, E.: Assessment of
  myocardial ischemia and viability using cardiac magnetic resonance. Curr
  Heart Fail Rep  \textbf{6}(3),  142--153 (2009)

\bibitem{Bonow_2011}
Bonow, R., et~al: Myocardial viability and survival in ischemic left
  ventricular dysfunction. N Engl J Med  \textbf{364}(17),  1617--25 (2011)

\bibitem{Gerber_2012}
Gerber, B., et~al: Prognostic value of myocardial viability by delayed-enhanced
  magnetic resonance in patients with coronary artery disease and low ejection
  fraction: impact of revascularization therapy. J Am Coll Cardiol
  \textbf{59}(9),  825--35 (2012)

\bibitem{Kerfoot2018}
Kerfoot, E., et~al: Automated {CNN}-based reconstruction of short-axis cardiac
  {MR} sequence from real-time image data. In: Image Analysis for Moving Organ,
  Breast, and Thoracic Images, 32 - 41. Springer (2018)

\bibitem{Kim_2000}
Kim, R., et~al: The use of contrast-enhanced magnetic resonance imaging to
  identify reversible myocardial dysfunction. N Engl J Med  \textbf{343}(20),
  1445--53 (2000)

\bibitem{Kramer_2015}
Kramer, C., et~al: Role of cardiac {MR} imaging in cardiomyopathies. J Nucl Med
   \textbf{56},  39S--45S (2015)

\bibitem{Lalande2020}
Lalande, A., et~al: {Emidec: A Database Usable for the Automatic Evaluation of
  Myocardial Infarction from Delayed-Enhancement Cardiac MRI}. Data
  \textbf{5}(4), ~89 (2020)

\bibitem{Lee_2020}
Lee, E., et~al: Practical guide to evaluating myocardial disease by cardiac
  {MRI}. Am J Roentgenol  \textbf{214}(3),  546--556 (2020)

\bibitem{Leiner_2019}
Leiner, T., et~al: Machine learning in cardiovascular magnetic resonance: basic
  concepts and applications. J Cardiovasc Magn Reson  \textbf{21}(1), ~61
  (2019)

\bibitem{loureno2020left}
Lourenço, A., et~al: Left atrial ejection fraction estimation using {SEGAN}et
  for fully automated segmentation of {CINE} {MRI} (2020)

\bibitem{Mantilla_2015}
Mantilla, J., et~al: Detection of fibrosis in late gadolinium enhancement
  cardiac {MRI} using kernel dictionary learning-based clustering. In:
  Computing in Cardiology Conference (CinC). pp. 357--360 (2015)

\bibitem{Narula_2016}
Narula, S., et~al: Machine-learning algorithms to automate morphological and
  functional assessments in {2D} echocardiography. J Am Coll Cardiol
  \textbf{68}(21),  2287--95 (2016)

\bibitem{Soriano_2005}
Soriano, C., et~al: Noninvasive diagnosis of coronary artery disease in
  patients with heart failure and systolic dysfunction of uncertain etiology
  using late gadolinium-enhanced cardiovascular magnetic resonance. J Am Coll
  Cardiol  \textbf{45}(5),  743--48 (2005)

\bibitem{sudre2017}
Sudre, C., et~al: Generalised {D}ice overlap as a deep learning loss function
  for highly unbalanced segmentations. In: Deep Learning in Medical Image
  Analysis and Multimodal Learning for Clinical Decision Support. pp. 240--248.
  Springer International Publishing, Cham (2017)

\bibitem{Weinsaft_2007}
Weinsaft, J., Klem, I., Judd, R.: {MRI} for the assessment of myocardial
  viability. Cardiol Clin  \textbf{25}(1),  35--36 (2007)

\end{thebibliography}
\bibliographystyle{splncs04.bst}

\end{document}